\documentclass[12pt]{article}
\usepackage{latexsym}
\usepackage{amsmath}
\usepackage{amssymb}
\catcode`\@=11
\newcount\hour
\newcount\minute
\newtoks\amorpm \hour=\time\divide\hour by 60\minute
=\time{\multiply\hour by 60 \global\advance\minute by-\hour}
\edef\standardtime{{\ifnum\hour<12 \global\amorpm={am}%
        \else\global\amorpm={pm}\advance\hour by-12 \fi
        \ifnum\hour=0 \hour=12 \fi
        \number\hour:\ifnum\minute<10
        0\fi\number\minute\the\amorpm}}
\edef\militarytime{\number\hour:\ifnum\minute<10
0\fi\number\minute}
\def\draftlabel#1{{\@bsphack\if@filesw {\let\thepage\relax
   \xdef\@gtempa{\write\@auxout{\string
      \newlabel{#1}{{\@currentlabel}{\thepage}}}}}\@gtempa
   \if@nobreak \ifvmode\nobreak\fi\fi\fi\@esphack}
        \gdef\@eqnlabel{#1}}
\def\@eqnlabel{}
\def\@vacuum{}
\def\marginnote#1{}
\def\draftmarginnote#1{\marginpar{\raggedright\scriptsize\tt#1}}
\def\draft{
        \pagestyle{plain}
        \overfullrule=2pt
        \oddsidemargin -.1truein
        \def\@oddhead{\sl \phantom{\today\quad\militarytime} \hfil
        \smash{\Large\sl DRAFT} \hfil \today\quad\militarytime}
        \let\@evenhead\@oddhead
        \let\label=\draftlabel
        \let\marginnote=\draftmarginnote
        \def\ps@empty{\let\@mkboth\@gobbletwo
        \def\@oddfoot{\hfil \smash{\Large\sl DRAFT} \hfil}
        \let\@evenfoot\@oddhead}
        \def\@eqnnum{(\theequation)\rlap{\kern\marginparsep\tt\@eqnlabel}%
        \global\let\@eqnlabel\@vacuum}  }

\renewcommand{\theequation}{\thesection.\arabic{equation}}
\renewcommand{\thefootnote}{\fnsymbol{footnote}}

\def\appendix#1{\addtocounter{section}{1}\setcounter{equation}{0}
\renewcommand{\thesection}{\Alph{section}}
\section*{Appendix \thesection\protect\indent \parbox[t]{11.15cm}{#1}}
\addcontentsline{toc}{section}{Appendix \thesection\ \ \ #1}}

\def \la {\label}

\def\be{\begin{equation}}
\def\ee{\end{equation}}

\catcode`\@=11

\def\bea{\begin{eqnarray}}
\def\eea{\end{eqnarray}}
\def\beann{\begin{eqnarray*}}
\def\eeann{\end{eqnarray*}}
\def\beq{\begin{equation}}
\def\eeq{\end{equation}}
\def\ba{\begin{array}}
\def\ea{\end{array}}
\def\ben{\begin{enumerate}}
\def\een{\end{enumerate}}

 \def \la {\label}
 \def\be{\begin{equation}}
\def\ee{\end{equation}}

\def \la {\label}

\font\mybb=msbm10 at 11pt

\def\bb#1{\hbox{\mybb#1}}

\def\bR {\bb{R}}

\def \ee {\epsilon}

\def\be{\begin{equation}}
\def\ee{\end{equation}}

\def \la{\label}

\newcommand{\tB}{\text{\tiny $B$}}
\newcommand{\tA}{\text{\tiny $A$}}
\newcommand{\tC}{\text{\tiny $C$}}

\begin{document}
\date{April 2006}
\begin{titlepage}
\begin{center}
\hfill
{}
\vspace{3.5cm}

{\Large \bf  On the structure of k-Lie algebras }
\\[.2cm]

\vspace{2.5cm} {\large   G. Papadopoulos}

 \vspace{0.5cm}
Department of Mathematics\\
King's College London\\
Strand\\
London WC2R 2LS, UK\\

\end{center}

\vspace{3.5cm}
\begin{abstract}

We show that the structure constants of  $k$-Lie algebras, $k>3$, with a positive definite metric are the sum
of the volume forms of orthogonal $k$-planes. This generalizes the result for $k=3$ in arXiv:0804.2662  and arXiv:0804.3078, and
 confirms a conjecture in math/0211170.

\end{abstract}
\end{titlepage}
\newpage
\setcounter{page}{1}
\renewcommand{\thefootnote}{\arabic{footnote}}
\setcounter{footnote}{0}

\setcounter{section}{0}
\setcounter{subsection}{0}

Metric $k$-Lie algebras, $k>2$, have emerged in the investigation of maximally supersymmetric
supergravity solutions in \cite{josegeorge1}. In particular, one finds that the Killing spinor equations
of IIB supergravity require that the 5-form field strength of maximally supersymmetric backgrounds
to obey the Jacobi identity of  the structure constants of a 10-dimensional 4-Lie algebra. So the classification of such backgrounds
relies on the understanding of the solutions of the associated Jacobi identity. This was achieved in \cite{josegeorge2},
after a lengthy computation, for the particular case that applies in IIB supergravity and some other related cases. In the same paper, it was realized
that $k$-Lie algebras, $k>2$, are highly constrained. In particular for all cases investigated it was found that
the structure constants are the sum of the volume forms of orthogonal planes. So it was conjectured that this
is likely to be the case for all metric $k$-Lie algebras, $k>2$, with Euclidean  signature metrics.

More recently, $3$-Lie algebras have appeared in an attempt to construct a multiple M2-brane theory \cite{bl1, gustavsson, bl2}.
This  followed earlier attempts to construct
superconformal ${\cal N}=8$ Chern-Simons
 \cite{schwarz} and multiple M2-brane theories \cite{basu}. Some other aspects have been examined
in \cite{berman, raam, morozov, ho, gomis, bergshoeff}. Consistency
requires that one should be able to relate such a M2-brane theory  to the $U(N)$ maximally
supersymmetric gauge theory in 3-dimensions which describes $N$ D2-branes \cite{mp1, lambert, mp2, gran}.
This relation also implies that there must be 3-Lie algebras which contain the Lie algebra $\mathfrak{u}(N)$
of the gauge group of D2-branes.  This again raises the issue of the solution of the Jacobi identities for metric 3-Lie algebras
and requires that there must be more general solutions than those suggested in \cite{josegeorge2}.
However, it was shown in \cite{george} that one cannot embed most semi-simple Lie algebras in metric 3-Lie algebras. Moreover it was confirmed
in \cite{george, jan}
that the only solutions of the Jacobi identity of metric 3-Lie algebras with a positive definite metric are those stated
in the conjecture of \cite{josegeorge2}.

In this paper, we shall confirm the conjecture of \cite{josegeorge2} for a $k$-Lie algebra,  $\mathfrak{a}_{[k]}$,  with a positive definite metric.
In particular, we shall show that the structure constants of such an algebra are of the form
\bea
F=\sum_r\mu^r\,  d{\rm vol}(V_r)~,~~~~ V_r\subset \mathfrak{a}_{[k]}
\la{main}
\eea
where $V_r$ are $k$-planes,   $V_r\perp V_{r'}$   for $r\not= r'$ and $\mu^r$ are constants. The proof
of (\ref{main}) follows closely that of \cite{george} for 3-Lie algebras.

The Jacobi identity of a metric k-Lie algebra, $\mathfrak{a}_{[k]}$, is
\bea
F_{\tC[\tA_1\dots \tA_k} F^\tC{}_{\tB_1]\tB_2\dots \tB_k}=0~,~~~A,B,C,\dots=0,\dots, n-1~,
\la{jac}
\eea
where $F$ are the structure constants  and $n$ is the dimension of $\mathfrak{a}_{[3]}$ , respectively.
Compatibility with a metric requires that the structure constants $F$  are skew-symmetric in all indices\footnote{We raise and lower
indices with  the compatible metric and  do not distinguish between a vector space and its dual.}.
For $k=2$,  (\ref{jac}) is
the Jacobi identity of a standard metric Lie algebra.

To prove our result, first observe that
given a vector $X$ in $\mathfrak{a}_{[k]}$, one can associate a metric (k-1)-Lie algebra  $\mathfrak{a}_{[k-1]}(X)$ to  $\mathfrak{a}_{[k]}$ defined
as the orthogonal complement of $X$ in $\mathfrak{a}_{[k]}$ with structure constants $i_XF$. It is easy to verify
that $i_XF$ satisfies the Jacobi identity using (\ref{jac}). This allows us to prove the result inductively. In particular,
the validity of the conjecture has been confirm for $k=3$. Assuming next that it is valid for metric $(k-1)$-Lie algebras, we shall
show that it is also valid metric for $k$-Lie algebras.

To continue, without loss of generality, take   the vector field $X$ to be along the $0$ direction. Then
   split the indices as $A=(0, i)$, $B=(0,j)$ and so on, with $i,j, \dots=1, \dots, n-1$. Setting $A_k=B_k=0$ and the rest of the
free indices in the range $1, \dots, n-1$ in (\ref{jac}), it is easy to see that
\bea
f_{i_1\dots i_k}=F_{0i_1\dots i_k}
\eea
satisfy the Jacobi identity of  $(k-1)$-Lie algebras and $f$ are the structure constants of  $\mathfrak{a}_{[k-1]}(X)$. Thus
we have written $F$ as
\bea
F={1\over k!} f_{i_1\dots i_k}\,\,  e^0\wedge  e^{i_1} \wedge \cdots \wedge e^{i_k}+ {1\over (k+1)!} \phi_{j_1\dots j_{k+1}}\, e^{j_1} \wedge \cdots
 \wedge e^{j_{k+1}}
\la{0phi}
\eea
where $(e^0, e^i)$, $i=1,\dots, n-1$, is an orthonormal basis.

Next set $B_k=0$ and the rest of the free indices in the range $1, \dots, n-1$ in (\ref{jac}). Using the skew-symmetry of $F$, one finds that
\bea
\phi_{h[j_1\dots j_k} f^h{}_{j_{k+1}]i_1\dots i_{k-2}}=0~.
\la{invphi}
\eea
This implies that the (k+1)-form $\phi$ is invariant with respect to $\mathfrak{a}_{[k-1]}(X)$.

To proceed, we must describe the invariant forms $\phi$. For this, we first use the inductive hypothesis to write
the structure constants of  $\mathfrak{a}_{[k-1]}(X)$ as the sum of the volume forms of orthogonal $k$-planes, $U_r$. In particular,
without loss of generality, one has
\bea
f=\mu^1\, e^1\wedge\cdots \wedge e^k+ \mu^2\, e^{k+1}\wedge \cdots \wedge e^{2k}+\dots~= \sum_r\,\mu^r\, f_r,
\eea
where $\mu^1, \mu^2,\dots$ are constants. Next, observe that (\ref{invphi}) can be interpreted as an invariance
condition for $\phi$ under all $k!/3! (k-3)!$ choices of $\mathfrak{su}(2)$ algebras that can be embedded in each
orthogonal $k$-plane $U_r$. This is done by choosing 3 out of $k$ orthogonal directions in $U_r$ with respect to an orthonormal basis. In fact, this
invariance holds for every choice of an orthonormal basis in $U_r$. All these infinitesimal transformations generate $\mathfrak{so}(U_r)$. Thus,
the only invariant forms are the volume
forms of the $U_r$ planes and those forms that are along the directions $\mathfrak{m}$ in $\mathfrak{a}_{[k]}$ which are orthogonal
to both $e^0$ and $\mathfrak{a}_{[k-1]}(X)$, ie
\bea
\phi= \sum_{I,r}  \nu_I{}^r \rho^I\wedge  f_r+\xi~,~~~~\mathfrak{a}_{[k]}=e^0\wedge \mathfrak{a}_{[k-1]}(X)\oplus \mathfrak{m}~,
\eea
where $\nu_I{}^r$ are constants,  and  $\rho^I\in \Lambda^1(\mathfrak{m})$ and  $\xi\in \Lambda^{k+1}(\mathfrak{m})$.
Thus  $F$ can be rewritten as
\bea
F=\sum_{r} \sigma^r\wedge f_r+ \xi~,
\eea
for some constants $\mu^r\not=0$ and $\nu_I{}^r$, where
\bea
\sigma^r=  \mu^r e^0+ \sum_I \nu_I{}^r \rho^I~.
\eea
Using that $f_r$ and $f_{r'}$,  for $r\not= r'$, are mutually orthogonal, the Jacobi identity (\ref{jac}) implies that
$\sigma^r$ and $\sigma^{r'}$ are mutually orthogonal as well. Thus there is an orthogonal transformation in $\mathfrak{m}$ such that one can write
\bea
F=\sum_{r} \lambda_r\,\, e^r\wedge f_r+\xi~,
\eea
for some constants $\lambda_r$,
where $e^r$ belong to an orthonormal basis in $\mathfrak{m}$. In particular, one has
 $e^r\perp e^{r'}$ for $r\not=r'$ and
 $i_{e^r} f_s=0$ for all $r$ and $s$.

Furthermore, using the orthogonality of $f_r$ and $\xi$ and the Jacobi identity (\ref{jac}), one finds that
\bea
i_{e^r}\xi=0~.
\la{fin}
\eea
This can be easily seen from (\ref{jac}) by setting
$A_1,\dots A_k$ to take values in the $\mathfrak{a}_{[k-1]}(X)$  and $B_1, \dots, B_{k}$ to take values in $\mathfrak{m}$.
A consequence of (\ref{fin}) is that the (k+1)-form $\xi$ on $\mathfrak{m}$ satisfies (\ref{jac}),
ie $\mathfrak{m}$ is also a metric k-Lie algebra, $\mathfrak{b}_{[k]}$,  with structure constants $\xi$.
Since the dimension of $ \mathfrak{b}_{[k]}\subset \mathfrak{a}_{[k]}$  is strictly less than that of the original metric
k-Lie algebra $\mathfrak{a}_{[k]}$, the analysis can be repeated and it will terminate after a finite number of steps.
Thus, we have established the conjecture of (\cite{josegeorge2}) for all $k$-Lie algebras with positive definite metrics.
It is likely that the above proof can be adapted to show that the conjecture holds for Lorentzian signature metrics as well.

To interpret (\ref{jac})  in the context of Pl\"ucker-type of relations, define $MGr_{k+1}(n)$  as the sequences
of orthogonal  $(k+1)$-planes in a n-dimensional Euclidean vector space $\bR^n$.  This space can be thought of as a multiple Grassmannian.
We have now established that (\ref{jac}) describes an embedding of $MGr_{k+1}(n)$, $k>2$,
 in the projective space $P(\Lambda^k(\bR^n))$
and it should be investigated further.

\vskip 0.2cm

{\bf Note added:} A different proof of the described result has been given in \cite{nagy} using another approach.

\vskip 0.2cm

\section*{Acknowledgements}
 I would like to thank U Gran for many helpful discussions and comments.

 \setcounter{section}{0}

\end{document}